\documentclass[a4paper,11pt]{article}

\usepackage{jcappub} 

\usepackage[T1]{fontenc} 
\newcommand {\dd}{\mathrm{d}}

\title{\boldmath Localization of $U(1)$ gauge field by non-minimal coupling with gravity in Braneworlds}

\author[a,1]{Zhen-Hua Zhao,\note{Corresponding author.}}
\author[b]{Qun-Ying Xie,}
\author[c]{Chun-E. Fu,}
\author[d]{and Xiang-Nan Zhou }

\affiliation[a]{Department of Applied Physics,    Shandong University of Science and Technology,    Qingdao 266590, China}
\affiliation[b]{School of Information Science and Engineering,  Lanzhou University, Lanzhou 730000, China}
\affiliation[c]{Institute of Theoretical Physics, School of Physics, Xi’an Jiaotong University, Xi’an 710049, China}
\affiliation[d]{College of Physics and Information Engineering, Shanxi Normal University, Taiyuan 030000, China}
\emailAdd{zhaozhh78@sdust.edu.cn}

\abstract{

In this paper, we investigate the localization of the $U(1)$ gauge field on Randall-Sundrum-like braneworld models.
The localization of the $U(1)$ gauge field is important because it plays a fundamental role in the branworld theories.
To achieve the localization, we propose a  novel action with a   non-minimal coupling between the $U(1)$  gauge field and  gravity. 
We find that the mass spectrum of the gauge field is continuous, without any gap between the zero-mass mode and the massive modes, and  except for the zero-mass mode all the massive modes are not localized on the brane. 
Furthermore, the massive modes have negative squared masses, indicating they are tachyonic. 
Our analysis can be applied to a wide range of thin and thick braneworld scenarios, provided that the five-dimensional spacetime is asymptotically anti-de Sitter.

 }

\begin{document}
\maketitle
\flushbottom

\section{Introduction }
In order to solve the hierarchy problem, Randall and Sundrum proposed a new branworld theory (RSI) with a compact extra-dimension \cite{Randall199983}. 
Further, they argued for a non-compact extra-dimensional model that can reproduce the familiar four-dimensional gravity \cite{Randall199983a}  (RSII). 
In RSII, the $Z_{2}$ symmetry is held along the infinite extradimension,  the 5D bulk is anti-de Sitter (AdS$_{5}$), and the brane is flat.
Gremm extended the RSII model to a thick braneworld one \cite{Gremm2000478}, in which the bulk is not anti-de Sitter, but the AdS$_{5}$ spacetime is monotonically restored as the extra dimension approaches infinity.
From here on, we refer to models with similar geometry to those in Refs. \cite{Randall199983a, Gremm2000478} as RSII-like models.

In RSII-like models with an infinite extradimension, matter fields must be localized to ensure that the action of 4D is finite after dimensional reduction. 
At least the zero-mass mode should be localized. 
The localization of gravitational \cite{Randall199983a,Gremm2000478}, scalar \cite{Bajc2000}, and fermion \cite{Rubakov1983,Randjbar-Daemi2000,Ringeval200265,Koley200522,Melfo2006,LiuZhangZhangDuan2008,Liu2018} fields can be easily realized without complex coupling terms that exceed the Standard Model.

But the localization of the gauge fields, { especially the non-Abelian gauge fields \cite{Oda2001a,BatellGherghetta2007,OhtaSakai2010,OhtaSakai2012,AraiBlaschkeEtoSakai2013a,
AlencarLandimMunizFilho2015,AraiBlaschkeEtoSakai2017,
AraiBlaschkeEtoSakai2018,AraiBlaschkeEto2021}, is more difficult. This paper will focus on the  $U(1)$ gauge field. The standard action of $U(1)$ gauge field,}
\begin{equation}
S=-\frac{1}{4}\int \dd^5x\sqrt{-g} \mathcal{F}_{A B} \mathcal{F}^{A B}, \label{action0}
\end{equation}
can not result the localization \cite{Pomarol2000} in RSII-like braneworld models, 
where $\mathcal{F}_{A B} =\partial_A \mathcal{A}_B-\partial_B \mathcal{A}_A$ is the field strength.

In order to discuss the localization of the  $U(1)$ gauge field in the RSII-like model, there are usually two types of approaches that can be adopted. One is to consider the non-minimal coupling of the vector and gravitational fields  \cite{Ghoroku2002,GolovnevMukhanovVanchurin2008,Germani2012,ZhaoXieZhong2014,
 AlencarLandimTahimCosta2014,AlencarLandimMunizFilho2015,Vaquera-AraujoCorradini2015,ZhaoXie2017,FreitasAlencarLandim2019,SuiGuoXieLiu2020,
 deOliveiraJuniorTahimAlencarEtAl2020}, and the other is to consider the coupling of the vector and scalar fields  \cite{IsozumiOhashiSakai2003,ChumbesHoffHott2012,ZhaoLiuZhong2014,Kehagias2001504,
 SuiZhao2017,AraiBlaschkeEtoEtAl2019,EtoKawaguchi2019,AraiBlaschkeEto2021,FuZhaoSun2022}.


%

The non-minimal coupling of the gravitational field and the U(1) gauge field has also been studied in four-dimensional spacetime \cite{Prasanna1971,Buchdahl1979,Mueller-Hoissen1988}.  
In the braneworld scenario, Germani proposes a non-minimal coupling mechanism with the following gauge field action \cite{Germani2012}
 \begin{equation}
S_v=-\frac{1}{4e^2_5}\int \dd^4 x \dd y \sqrt{-g} \left[\mathcal{F}_{A B} \mathcal{F}^{ A B}+ \frac{1}{M_v^2}\Delta^{A B}_{v\ \ \  C D} \mathcal{F}_{A B} \mathcal{F}^{C D}\right], \label{GerA}
\end{equation}
where 
\begin{equation}
\Delta^{A B}_{v\ \ \ C D} \equiv \frac{1}{8}\mathcal{R}^{A B}_{\ \ \ C D} -\frac{1}{2}\mathcal{R}^{[ A}_{\ \ [C}
\delta^{B]}_{\ D]} +\frac{1}{8}\mathcal{R}\, \delta^{[A }_{\ [C}\delta^{B]}_{\ D]}.
\end{equation}
Furthermore, it can be shown that
\[ \Delta^{A B}_{v\ \ \ C D} =- 8  \mathcal{L}^{A B}_{\  \ \ C D}, \]
where $ \mathcal{L}^{A B}_{\  \ \ C D} $  is just the  double dual of Riemann tensor \cite{MisnerThorneWheeler1973,ChowdhuryRahmanSabuj2018}. The work of Germani can be seen as an improvement of the work of Dvali, Gabadadze and Shifman \cite{Dvali2001}. But they \cite{Germani2012,Dvali2001} all do not obtain a localised solution. Based on the work of Germani \cite{Germani2012}, Alencar et al. \cite{AlencarLandimMunizFilho2015} added a mass term of the gauge field to the action \eqref{GerA} in order to achieve a localised solution of the $U(n)$ gauge field. However, the gauge symmetry is again broken by the addition of the mass term.



%
%

In this paper, based on the work of Germani \cite{Germani2012}, we introduce a new non-minimal coupling mechanism that can not only produce a localised zero-mass mode solution for the $U(1)$ gauge field, but also preserve the gauge symmetry of the action.

This paper is organized as follows. 
We review our mechanism in section \ref{sec1}. The localization of zero-mass and massive modes is discussed in sections \ref{sec3} and \ref{sec4}, respectively. 
We demonstrate the effectiveness of our analysis with a test using a $f(R)$ braneworld model in section \ref{sec5}.  
Finally, we present our conclusions and discussions in section \ref{Cons}.

\section{Localization Mechanism \label{sec1}}

The line element of the 5D spacetime is assumed to be 
\begin{eqnarray}
\dd s^2={g}_{A B}\dd x^A dx^B = e^{2\zeta(y)}\eta_{\mu \nu}\dd x^{\mu}\dd x^{\nu}+\dd y^2,\label{metric}
\end{eqnarray}
where the bulk indices $A,B,\cdots$ run as $\{$0, 1, 2, 3, 4$\}$, the brane indices $\mu, \nu,\cdots$ run as $\{$0, 1, 2, 3$\}$,  $\eta_{\mu \nu}=\text{diag}{(-1,1,1,1)}$ is the metric on the branes, and $e^{2 \zeta(y)}$ is the warp factor. For an AdS$_5$ spacetime or an asymptotic AdS$_5$ spacetime, when $y$ tends to infinity the limit solution of $\zeta(y)$ is \cite{Randall199983a}
\begin{equation} 
\zeta(y\to \pm \infty) = - k |y|.  \label{alphainf}
\end{equation}

The conventions for the Christoffel connection and the Riemann tensor follow those in ref. \cite{MisnerThorneWheeler1973}.
The Christoffel connection is
\begin{equation}
\Gamma^{A}_{\ \ B C} = \frac{1}{2} g^{A D} ( 
                                   \partial_{C}{g_{B D}} 
                                  +\partial_{B}{g_{C D}}
                                  -\partial_{D}{g_{B C}} )
\end{equation}
 and the Riemann tensor 
\begin{equation}
\mathcal{R}^{C}_{\ \ D A B } = \partial_{A}{{\Gamma}^{C}_{\ \ B D}} 
-\partial_{B}{\Gamma^{C}_{\ \ A D}}
+\Gamma^{C}_{\ \ A N} \Gamma^{N}_{\ \ B D}
-\Gamma^{C}_{\ \ B N} \Gamma^{N}_{\ \ A D}.
\end{equation}
The double dual of the Riemann tensor is \cite{MisnerThorneWheeler1973,ChowdhuryRahmanSabuj2018}
\begin{eqnarray}
{\mathcal{L}}^{A B}_{\ \ \ C D}  \equiv  -\mathcal{R}^{A B}_{\ \ \ C D} +4\mathcal{R}^{[ A}_{\ \ [C}
\delta^{B]}_{\ D]} - \mathcal{R}\, \delta^{[A }_{\ [C}\delta^{B]}_{\ D]}.
\end{eqnarray}

In this scenario, the gravitational field is taken as the background, and the back reaction from the $U(1)$ gauge field is neglected. The action of the 5D $U(1)$ gauge field we proposed is
\begin{equation}
S=-\frac{1}{4}\int \dd^5 x \sqrt{-g}(\gamma_1 \mathcal{R}^{A B}_{\ \ \ C D}   +\gamma_2 {\mathcal{L}}^{A B}_{\ \ \ C D}) \mathcal{F}_{A B} \mathcal{F}^{ C D}\label{action4},
\end{equation}
where $\gamma_1$ and $\gamma_2$ are coupling constants, and
\begin{equation}\mathcal{R}^{A B}_{\ \ \ CD} = g^{B N} \mathcal{R}^{A}_{\ \ N C D}.  \end{equation}
Compared to the action of Germani \eqref{GerA}, the main change we made is that we dropped the term $\mathcal{F}_{A B} \mathcal{F}^{ A B}$ and added a term $\gamma_1 \mathcal{R}^{A B}_{\ \ \ C D} \mathcal{F}_{A B} \mathcal{F}^{ C D}$.


With the separation of the variables  $\mathcal{A}_{A}= a_{A}(x)\phi(y)$  and  the gauge condition $\partial_{\mu} \mathcal{A}^{\mu}=0$, $\mathcal{A}_4=0$, after a long but straightforward derivation, the  action (\ref{action4}) is reduced to
\begin{equation}
S=-\frac{1}{4}\int \dd y F_1 (y) \phi(y)^2\int \dd^4x(f_{\mu \nu}f^{\mu \nu}+2 m^2 a_{\mu} a^{\mu})\label{action5},
\end{equation}
where $f_{\mu\nu}=\partial_{\mu}a_{\nu}-\partial_{\nu}a_{\mu}$ is the 4D gauge field tensor, $m$ is the mass of 4D gauge field, and 
\begin{equation}
F_1(y)=(6 \gamma_1 -2 \gamma_2)\zeta'(y)^2 +4 \gamma_1 \zeta''(y). \label{F1}
\end{equation}
 At the same time, $\phi(y)$ satisfies the following equation
\begin{equation}
-\partial_y (F_2 \phi' e^{2\zeta(y)} )=F_1 \phi m^2 \label{eq},
\end{equation}
 where
 \begin{equation}
F_2(y)=(6 \gamma_1 -2 \gamma_2)\zeta'(y)^2 -2 \gamma_1 \zeta''(y). \label{F2}
\end{equation}
The boundary conditions of eq.\eqref{eq} can be either the Neumann $\phi'(\pm \infty)=0$ or the Dirichlet $\phi(\pm \infty)=0$ \cite{Gherghetta2010}.

From eq.\eqref{action5}, one can find that in order to recover the conventional 4D action  the following integration should be finite, at least for the zero-mass mode,
\begin{equation}
I\equiv\int_{-\infty}^{+\infty} \dd y F_1 (y)\phi^2(y) . \label{int}
\end{equation}

\section{Zero-Mass Mode at $y$ Coordinate \label{sec3}}

For the zero-mass mode, $m=0$, eq. (\ref{eq}) is reduced to
\begin{equation}
\partial_y (F_2(y) \phi' e^{2\zeta} )=0. \label{eq0}
\end{equation}
The general solution is
\begin{equation}
\phi_0=\mathrm{d}_0 +\mathrm{d}_1 \int \frac{e^{-2\zeta}}{F_2 (y)} \dd y, \label{appzeroso1} 
\end{equation} 
 $\mathrm{d}_0$ and $\mathrm{d}_1$ are integration constants. Since the 5D spacetime holds the $\mathbb{Z}_2$ symmetry about the coordinate $y$, $\zeta(y)$ and $F_2 (y)$ are even functions of $y$, so $\phi(y)$ is the odd one. The Dirichlet boundary conditions $\phi_n(\pm \infty)=0$ lead to $\mathrm{d}_0=0$ and $\mathrm{d}_1=0$. But the Neumann boundary conditions $\phi_n'(\pm \infty)=0$ lead only to $\mathrm{d}_1=0$, so the unique zero mass mode solution is
 \begin{equation}
\phi=\mathrm{d}_0 .\label{sphi0}
\end{equation}

Substituting the above solution \eqref{sphi0} into the integration \eqref{int}, we obtained
\begin{eqnarray}
\int_{-\infty}^{+\infty} \dd y F_1(y)\phi_0^2(y)=\mathrm{d}_0 ^{2}\int_{-\infty}^{+\infty}  F_1(y)\, \dd y \label{int2}.
\end{eqnarray}
Since $F_1(y)$ is continuous, the convergence of the integration is determined by the asymptotic behavior of $F_1(y)$ at infinity along the extra dimension. 
This requires that the asymptotic solution of  $F_1(y)$  at infinity takes the form as
\begin{equation}
F_1(y\to \pm \infty) \propto y^{-p} \label{Cd2}
\end{equation}
 with $p > 1$.

By using the $\mathbb{Z}_2$ symmetry,  we only need to discuss the case of $y\to + \infty$.
  The exact form of the $F_1$  is determined by the solution of $\zeta(y)$ \eqref{F1}. But we have only one asymptotic solution \eqref{alphainf} of $\zeta(y)$ at hand. In order to discuss the asymptotic behavior of $F_1(y)$ at infinity for the thick braneworld case, we qualitatively add a first-order infinitesimal term in $\zeta(y)$ \eqref{alphainf} as
\begin{equation}
\zeta(y \to + \infty) \sim - \,k\, y - \mathrm{C}_1 y^{-q},\label{alphainf2}
\end{equation}
where $q >0$. 
Substituting \eqref{alphainf2} into \eqref{F1} we gets 
\begin{eqnarray}
F_1 (y \to + \infty) &\sim& (6 \gamma_1-2 \gamma_2)(\mathrm{C}_1 q y^{-q-1} -k)^2  - 2 \gamma_1 (q+1)q y^{-q-2} \nonumber\\
 &\sim& (6 \gamma_1-2 \gamma_2) k^2 - 2 \gamma_1 (q+1)q y^{-q-2} .\label{chiinf}
\end{eqnarray}
Since $q+2>1$, the second term of the right side of the above equation \eqref{chiinf} satisfies the condition \eqref{Cd2}, so to finally reach the condition \eqref{Cd2}, the first term must go to zero, namely
\begin{equation}
6 \gamma_1-2 \gamma_2 = 0 \Rightarrow \gamma_2 = 3 \gamma_1 .\label{cd2}
\end{equation}
So $\gamma_2 = 3 \gamma_1 $ is the key condition for locating the zero-mass mode. And this conclusion is independent of the exact solution of $\zeta(y)$. At the same time,  by using the condition \eqref{cd2}, the functions $F_1(y)$ \eqref{F1} and $F_2(y)$ \eqref{F2} are simplified into
\begin{equation}
F_1(y)= 4 \gamma_1 \zeta''(y). \label{F11}
\end{equation}
\begin{equation}
F_2(y)= -2 \gamma_1 \zeta''(y)= -\frac{1}{2} F_1(y) . 
\end{equation}


\section{Massive  and Zero-mass Modes  at $z$ Coordinate } \label{sec4}


It is more convenient to discuss the massive modes at the $z$ coordinate, using the line element 
\begin{equation}
\dd s^2=e^{2 \zeta(z)}(\eta_{\mu \nu}\dd x^{\mu}\dd x^{\nu}+\dd z^2), \label{gcf}
\end{equation}
which is derived from the line element \eqref{metric} by the coordinate transformation
\begin{equation}\label{ytoz}
\left\{
\begin{array}{lll}
\dd z &=& e^{- \zeta(y)} \dd y ,\\
z &=& \int e^{- \zeta(y)} \dd y 
\end{array} \right.
\end{equation}
with the constraint $z(y=0)=0$.

By using the gauge choice $\mathcal{A}_4=0$ and the decomposition  
\begin{equation}
\mathcal{A}_{\mu}=a_{\mu}(x) \left(F_1(z) e^{\zeta(z)}\right)^{-1/2} \hat{\phi}(z).
\end{equation}
The action of the five-dimensional gauge field (\ref{action4})
is reduced to
\begin{equation}
S=-\frac{1}{4}\int \dd z \hat{\phi}^2(z)\int \dd^4x(f_{\mu\nu}f^{\mu\nu}+2 m^2 a_{\mu}a^{\mu})\label{action7},
\end{equation}
where $\hat{\phi}(z)$ satisfies the following Schr\" odinger-like equation
\begin{equation}
-\hat{\phi}''+V(z)\hat{\phi}=\hat{m}^2 \hat{\phi}. \label{eq3}
\end{equation}
The above equation takes the form of an analog of non-relativistic quantum mechanics, where 
\begin{equation}\label{eq32}
\hat{m}^2 = -2 m^2/3
\end{equation}
and $m$ is the mass of the 4D gauge field. The effective potential 
\begin{equation}
V(z)=K'(z)^2+K''(z)\label{Vz}
\end{equation}
with
\begin{equation}
K'(z)=\frac{1}{2}\left(\frac{F_1'(z)}{F_1(z)}+\zeta'(z)\right),
\end{equation}
and the form of the function $F_1$ is changed from \eqref{F11} to 
\begin{equation} F_1(z) = 4 \gamma_1( \zeta''(z) - \zeta'(z)^2) e^{-2 \zeta(z)}. 
\end{equation}

Equation \eqref{eq3} can be recast in a supersymmetric quantum mechanical form
\begin{equation}
\bigg(\frac{\dd}{\dd z}+K'(z)\bigg)\bigg(-\frac{\dd}{\dd z}+K'(z)\bigg)\hat{\phi}=\hat{m}^2 \hat{\phi}. \label{eq4}
\end{equation}
Which excludes the solutions of $\hat{m}^2<0$ \cite{Sukumar1985,Bazeia2004,BogdanosDimitriadisTamvakis2006}. 
This is the reason why we do the coordinate transformation \eqref{ytoz}.

But the  relation of mass square of 4D particle $m^2 $ and $\hat{m}^2$ is \eqref{eq32}, that
means 
\begin{equation}
m^2\leq 0,
\end{equation}
namely, except for the zero-mass mode, all the  massive modes are { tachyonic}. 


The possibility of the localization of massive modes is determined by the behavior of potential $V(z)$ \eqref{Vz} at infinity. For  the $\mathbb{Z}_2$ symmetry about the coordinate $z$, without losing generality, we    discuss only the case of $y\to + \infty$ about $V(z)$. 

By using the {  limit solution} $\zeta(y)$ \eqref{alphainf}  and the coordinate transformation  \eqref{ytoz},
we can obtain the  {  limit solution} of $\zeta(z) $  at infinity,
\begin{equation}
 \zeta(z \to \infty) \sim -\ln ( k\, z) -  \mathrm{C}_2 z^{-n},      \label{alphainfzasy} 
\end{equation}
where $\mathrm{C}_2$ and $n$ (>0) are model dependent constants. Substituting \eqref{alphainfzasy} into \eqref{Vz}, the asymptotic solution of $V(z)$  reads
\begin{eqnarray}
 V(z \to \infty) =&&
  \frac{z^{-2 (n+1)}}{4 \left( \mathrm{C}_2 n-(n-1) z^n\right)^2} \left[\mathrm{C}_2^4 n^4  -2 \mathrm{C}_2^3 n^3 (4 n+1) z^n  +\mathrm{C}_2^2 n^2 \left(15 n^2+4 n-4\right) z^{2 n}\right.  \nonumber\\
  && \left. -2 \mathrm{C}_2 n \left(5 n^3+n^2-5 n-1\right) z^{3 n}  +(n-1)^2 \left(n^2+4 n+3\right) z^{4 n}\right].\label{Vz2}
\end{eqnarray}
The limitation of $V(z)$ is 
\begin{equation}
\lim_{z \to +\infty} V(z) = 0 ,\label{Vzinfty}
\end{equation}
which is  $\mathrm{C}_2$ and $n$ independent. 

This  means that all tachyonic massive KK modes are not localized on branes, which is  parameters independent   in our mechanism, so  only the zero-mass mode is localized as the photon in our 4D world.

The solution of the zero-mass mode is easily obtained from \eqref{eq4}
{ 
\begin{equation}
\hat{\phi}_0(z)=e^{\zeta(z)/2}\sqrt{F_1(z)}. \label{zeroz1}
\end{equation}
}
Since
\begin{equation}
\lim_{z \to +\infty} F_1 (z) = 0,
\end{equation}
and the limit solution of $e^{\zeta(z)/2}$ is
\begin{equation}
\lim_{z \to +\infty} e^{\zeta(z)/2}= (k z)^{-1/2},
\end{equation}
so the square integrability of $\hat\phi_0(z)$ in \eqref{action7} is approved.

\section{Test by $f(\mathcal{R})$  Braneworld  Model} 
\label{sec5}

Here we choose the $f(\mathcal{R})$ pure geometric thick braneworld model ~\cite{ZhongLiu2016} to test our above analysis. Compared to the one used by Gremm \cite{Gremm2000478}, the $f(\mathcal{R})$ model does not need to introduce a background scalar field and has similar analytic solutions, and the corresponding gravitational zero-mass mode is localized on the brane, leading to the four-dimensional Newton's law.

\begin{figure}[b]
\includegraphics[width= 0.4\textwidth]{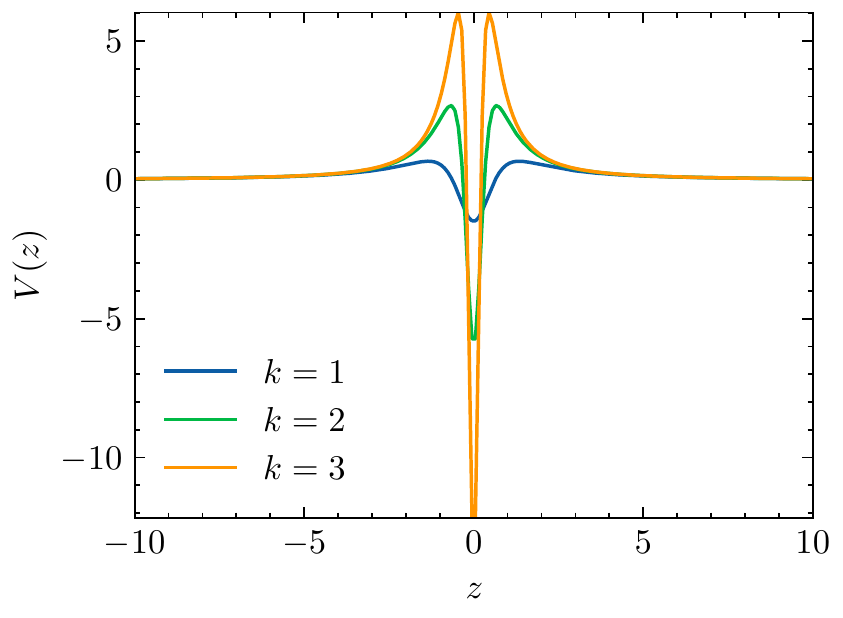}
\includegraphics[width= 0.4\textwidth]{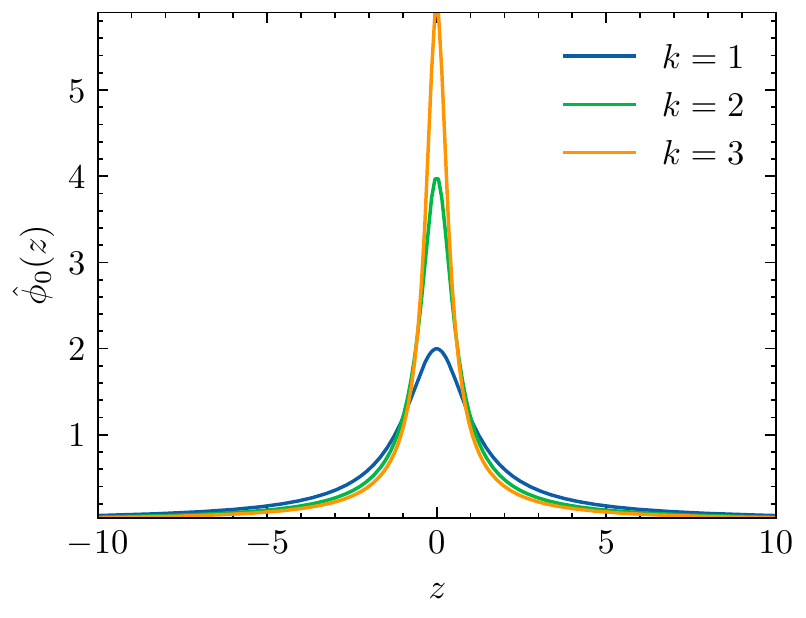}
\caption{ \label{fig1} The plots of the effective potential and zero-mass mode with parameter $\gamma_{1}= -1$ and different values of $k$. }
\end{figure}

The action of the $f(R)$  braneworld  model is
\begin{equation}
S=\frac{1}{4} \int \dd ^{4} x \dd y \sqrt{- g} f(R), \label{fR}
\end{equation} 
where
\begin{eqnarray}
\label{SolF1}
f(\mathcal{R})&=&\frac{4}{7} \left(6 k^2+\mathcal{R}\right)\cosh (\beta(w(\mathcal{R})) )  - \frac{2}{7} k^2 \sqrt{480-\frac{36 \mathcal{R}}{k^2}-\frac{3 \mathcal{R}^2}{k^4}} \sinh (\beta(w(\mathcal{\mathcal{R}})) ),
\end{eqnarray}
and
\begin{eqnarray}
\beta (w) &=& 2 \sqrt{3} \arctan \left(\tanh \left(\frac{w}{2} \right)\right), \nonumber\\ w(\mathcal{R})&=&\pm\text{arcsech}\left[\frac{\sqrt{20 +\mathcal{R}/k^2}}{2\sqrt{7}}\right].
\end{eqnarray}
The above action \eqref{fR} supports the following analytical solution of the warped factor 
\begin{equation}
\zeta (y)= - \ln (\cosh (k\, y)).
\end{equation}
After performing the coordinate transformation \eqref{ytoz}, the warped factor at the coordinate $z$ is changed into
\begin{eqnarray}
\zeta(z)&=&-\frac{1}{2} \log \left(k^2 z^2+1\right). \label{chiexam1}
\end{eqnarray}
Substituting the above analytic solution of $\zeta(z)$ \eqref{chiexam1} into the potential \eqref{Vz}, we get 
\begin{eqnarray}\label{Vz1}
V(z)=\frac{3 k^2 (-2 + 5 k^2 z^2)}{4 (1 + k^2 z^2)^2}, 
\end{eqnarray}
and
\begin{eqnarray}
\lim_{z\to +\infty} V(z) = 0 , \label{VI2}
 \end{eqnarray} 
which is consistent with eq. \eqref{Vzinfty}. The solution of the zero mass mode is
\begin{eqnarray}
\hat{\phi}_0 = 2 \sqrt{-\gamma_1} k \left(k^2 z^2+1\right)^{-3/4},
 \end{eqnarray} 
where $\gamma_1 <0$, and the exact value of $\gamma_1$ is determined by the normalization condition of $\hat{\phi}_0$. This zero-mass mode solution is localized. Some plots of $V(z)$ and $\hat{\phi}_0$ are shown in figure \ref{fig1}.

\section{Conclusions and Discussions} \label{Cons}

%
%
%

The localization of $U(1)$ gauge fields in Randall-Sundrum II (RSII)-like models has been a long-standing problem in braneworld theories. To address this issue, we propose a new non-minimal coupling action inspired by the work of Germani \cite{Germani2012}. Specifically, we modify the original action by dropping the term $\mathcal{F}_{A B} \mathcal{F}^{ A B}$ and adding a new term $\gamma_1 \mathcal{R}^{A B}_{\ \ \ C D} \mathcal{F}_{A B} \mathcal{F}^{ C D}$.

With our proposed action, we are able to obtain a localized zero-mass mode, which leads to the 4D conventional $U(1)$ gauge action in 4D.  Notably, our proposed action preserves gauge symmetry and avoids introducing new degrees of freedom into the scenario.

However, we also find that the massive KK modes exhibit  imaginary masses, which is a feature of the action and is independent of the parameters $\gamma_1$ and $\gamma_2$.

In summary, our proposed non-minimal coupling action offers a promising solution for the localization of the $U(1)$ gauge field in braneworlds. Nevertheless, the presence of tachyonic modes warrants further investigation into the stability of the theory.  It remains to be seen whether our proposed non-minimal coupling action will provide a viable solution for localizing other types of gauge fields in braneworld scenarios. 
In addition, while the Lagrangian \eqref{GerA} is proved to be unique in
four dimensions \cite{Horndeski1976},  is our action \eqref{action4} also unique in five-dimensional
spacetime? We would like to address this issue in a future work.

\section*{Acknowledgments}

{ This work is supported by the Scientific Research Foundation of Shandong University of Science and Technology for Recruited Talents (Grant No. 2013RCJJ026) and the Natural Science Foundation of Shaanxi Province (No. 2022JQ-037). We would like to thank Kasper Peeters for his computer algebra program Cadabra \cite{Peeters2007,Peeters2007a,Peeters2018} and  some of the calculations were done with  it.



 \bibliographystyle{JHEP}
 \bibliography{D:/yun/Zotero/library.bib}

\end{document}